%% file: 00_main.tex
\documentclass[conference]{IEEEtran}

\pdfoutput=1

\input{01_preamble.tex}
\input{02_acronym.tex}

\bibliography{bibliography.bib}

\begin{document}

\title{Achieving Determinism in Adaptive AUTOSAR}

\author{
  \IEEEauthorblockN{
    Christian Menard\IEEEauthorrefmark{1},
    Andrés Goens\IEEEauthorrefmark{1},
    Marten Lohstroh\IEEEauthorrefmark{2} and
    Jeronimo Castrillon\IEEEauthorrefmark{1}
  }
  \IEEEauthorblockA{\IEEEauthorrefmark{1}
    Center for Advancing Electronics Dresden (cfaed), TU Dresden, Dresden, Germany\\
    \{christian.menard, andres.goens, jeronimo.castrillon\}@tu-dresden.de
  }
  \IEEEauthorblockA{\IEEEauthorrefmark{2}
    Department of EECS, UC Berkeley, USA\\
    marten@berkeley.edu
  }}

\maketitle

\begin{abstract}
  \input{03_abstract.tex}
\end{abstract}

\begin{IEEEkeywords}
automotive engineering, reliability and testing, software and system safety,
software engineering
\end{IEEEkeywords}

\input{10_intro.tex}

\input{20_autosar.tex}
\input{30_discrete_events.tex}
\input{40_case_study.tex}
\input{50_related_work.tex}
\input{60_conclusion.tex}

\section*{Acknowledgment}
We thank Edward A. Lee for his feedback on an earlier version of this paper.
This work was supported in part by the KMU-innovativ project EM-RAM funded by
German Federal Ministry of Education and Research (BMBF). The work was further
supported in part by the National Science Foundation (NSF), award \#CNS-1836601
and the iCyPhy Research Center (Industrial Cyber-Physical Systems), supported
by Camozzi Industries, Denso, Ford, Siemens, and Toyota.

\printbibliography

\end{document}

%% file: 01_preamble.tex
\usepackage[utf8]{inputenc}
\usepackage[american]{babel}
\usepackage{blindtext}
\usepackage{enumitem}
\usepackage{pifont}
\usepackage{amsmath,amsfonts}
\usepackage{microtype}
\usepackage{algorithmic}
\usepackage{graphicx}
\usepackage{textcomp}
\usepackage{xcolor}
\usepackage{tikz}
\usepackage[frozencache]{minted}
\usepackage{csquotes}
\usepackage{moresize}
\usepackage{todonotes}
\usepackage[style=ieee,maxbibnames=6]{biblatex}
\usepackage[nolist]{acronym}
\usepackage[normalem]{ulem} %
\usepackage{siunitx}
\usepackage[scaled=0.86]{helvet}

\renewcommand*{\bibfont}{\footnotesize}

\definecolor{c1}{HTML}{FFFFD9}
\definecolor{c2}{HTML}{EDF8B1}
\definecolor{c3}{HTML}{C7E9B1}
\definecolor{c4}{HTML}{7FCDBB}
\definecolor{c5}{HTML}{41B6C4}
\definecolor{c6}{HTML}{1D91C0}
\definecolor{c7}{HTML}{225EA8}
\definecolor{c8}{HTML}{0C2C84}

\usetikzlibrary{positioning}
\usetikzlibrary{backgrounds}
\usetikzlibrary{fit}
\usetikzlibrary{calc}
\usetikzlibrary{arrows.meta}

\DeclareSIUnit\ms{\milli\second}

%% file: 02_acronym.tex
\begin{acronym}
 \acro{SoA}{service-oriented architecture}
 \acro{WCET}{worst-case execution time}
 \acro{CPS}{cyper-physical system}
 \acro{ECU}{engine control unit}
 \acro{CPS}{cyber physical system}
 \acro{EBA}{emergency brake assistant}
 \acro{APD}{Adaptive Platform Demonstrator}
 \acro{AP}{Adaptive Platform}
 \acro{CP}{Classic Platform}
 \acro{DE}{discrete event}
 \acro{MoC}{model of computation}
 \acro{SWC}{software component}
 \acro{LET}{logical execution time}
\end{acronym}

%% file: 03_abstract.tex
AUTOSAR \ac{AP} is an emerging industry standard that tackles the
challenges of modern automotive software design, but does not provide adequate
mechanisms to enforce deterministic execution.  This poses profound challenges
to testing and maintenance of the application software, which is particularly
problematic for safety-critical applications.  In this paper, we analyze the
problem of nondeterminism in \ac{AP} and propose a framework for the design of
deterministic automotive software that transparently integrates with the
\ac{AP} communication mechanisms. We illustrate our approach in a case study
based on the brake assistant demonstrator application that is provided by the
AUTOSAR consortium. We show that the original implementation is
nondeterministic and discuss a deterministic solution based
on our framework.

\acresetall

%% file: 10_intro.tex
\section{Introduction}
\label{sec:intro}

Designing and developing software for automotive applications is challenging
due to stringent safety and real-time requirements. New use cases like the
self-driving car have caused a dramatic increase in complexity and
computational demands of automotive software. The
AUTOSAR\footnote{\url{https://www.autosar.org/}} consortium addresses the
challenges in industrial automotive software design by standardizing the design
process, the runtime environment, and the common software framework. The
consortium maintains two standards called \emph{\acf{CP}} and \emph{\acf{AP}}
that serve different goals and requirements.  The former is already established
in industry and mostly intended for hard real-time applications with low
computational complexity deployed on single-core processors. The latter was
introduced more recently in order to handle applications with a high
computational demand and the need for maintaining ongoing interaction with a
changing environment.

One particular challenge that the automotive industry faces is nondeterminism
in the software architecture.  A deterministic program yields exactly one
behavior given an initial state and inputs, whereas a nondeterministic program
may yield many. Nondeterminism may be harmless in some applications while it
can lead to unintended and unanticipated behavior in others. In either case,
unintended or ``accidental'' nondeterminism tends to impair testability and
negatively impact maintainability of the software.  This serves as a compelling
argument for allowing nondeterminism only when
needed~\cite{lee2006-problem-with-threads}.  In safety-critical systems,
unintended system behavior could translate into physical damage, injury, or
even loss of life. For this reason, the nuclear, aeronautics, and railways
industries often rely on synchronous languages like
LUSTRE~\cite{halbwachs1991synchronous}, Esterel~\cite{berry1992esterel}, and
SCADE~\cite{scade-berry} to rule out nondeterminism in their designs of
safety-critical software~\cite{Boulanger:2015:SLA:2821094}.

\begin{figure}[t]
  \begin{minipage}{0.55\linewidth}
    \begin{minted}[fontsize=\footnotesize,autogobble]{c++}
      int main() {
        s = ServiceProxy();

        s.set_value(1);
        s.add(2);
        result = s.get_value();

        std::cout << result.get();
        return 0;
      }
    \end{minted}
  \end{minipage}\hfill
  \begin{minipage}{0.44\linewidth}
    \centering
    \input{plot/intro_dist.tex}
  \end{minipage}
  \caption{A nondeterministic AUTOSAR \acf{AP} client/server application. The
client manipulates the server's state variable in a series of (non-blocking) procedure calls.
The client prints out one of four different results, distributed as shown in the graph
on the right.}
\label{fig:intro_example}
\vspace{-0.5cm}
\end{figure}
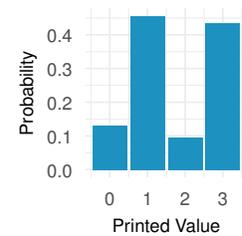

In the latest iteration of its well established \ac{CP} standard, the AUTOSAR
consortium introduced support for the \ac{LET}
paradigm~\cite{kirsch2012-logical-execution,autosar2018-specific-timing-extension}.
This can be used to build deterministic software while exploiting the
parallelism of multi-core
architectures~\cite{biondi2018-achievin-predictab}. However, in \ac{AP} no
effective means have been provided for ensuring deterministic execution. On the
contrary, \ac{AP} applications are commonly distributed, and the
service-oriented model of \ac{AP} poses major challenges to the
development of deterministic software.

Consider the C++ code in Figure~\ref{fig:intro_example} that implements a naive
client/server application in AUTOSAR \ac{AP}. At first glance, C++ being a
procedural language, the code suggests that the printed value should be {\tt 3}.
However, potentially unbeknownst to the programmer that wrote the client, the
server implements methods {\tt set\_value} and {\tt add} in a non-blocking
fashion. And while the server implementation enforces mutual exclusion between
the execution of method invocations, by default, the runtime environment maps each
invocation to a different thread~\cite{autosar2019-specific-communic-managemen}, meaning the order in which the calls are handled
is determined purely by the thread scheduler. As a result, no order is enforced
on the handling of calls to {\tt set\_value}, {\tt add}, and {\tt
get\_value}, leading to nondeterministic results.

Of course, the client could instead serialize each method call by waiting for
the future returned by the server to resolve prior to invoking the next method
call; and the server could inform the runtime to use a single thread rather
than multiple. However, multi-threading may be necessary to meet performance
requirements, yet it is often far from obvious how this may lead to
nondeterminism in realistic AUTOSAR applications which are, of course,
incomparably more complex than this simple example. Therefore, we argue that
the software designer should not be responsible for engineering solutions to
concurrency problems in order to achieve determinism. Rather, the underlying
model should allow for the exploitation of concurrency in ways that preserve
determinism, making it easy to write deterministic programs and requiring
explicit directions from the programmer to forgo determinism.

In this paper, we address the lack of an execution model capable of guaranteeing
determinism in AUTOSAR \ac{AP} and show how this can be solved.  We
make the following contributions:%
\begin{itemize}
\item We analyze AUTOSAR \ac{AP} and identify three potential sources of nondeterminism (Section~\ref{sec:autosar}).
\item We propose a solution based on reactors~\cite{lohstroh2019-invited}, a
recently introduced deterministic reactive programming model in which software
components are coordinated under a discrete-event semantics. We demonstrate that
reactors can integrate with the existing communication mechanisms of AUTOSAR
\ac{AP} and deliver determinism while maintaining compatibility with the
standard~(Section~\ref{sec:de}).
\item We present a case study based on the brake assistant application provided
by the \ac{APD}.  We show that this
application exhibits nondeterminism that directly translates into
problematic behavior. Finally, we describe a deterministic implementation based on
reactors that addresses the problem~(Section~\ref{sec:case-study}).
\end{itemize}

%% file: plot/intro_dist.tex
\begin{tikzpicture}[x=1pt,y=1pt,font=\sffamily]
\definecolor{fillColor}{RGB}{255,255,255}
\begin{scope}
\definecolor{drawColor}{gray}{0.92}

\path[draw=drawColor,line width= 0.3pt,line join=round] ( 30.87, 37.39) --
	( 92.06, 37.39);

\path[draw=drawColor,line width= 0.3pt,line join=round] ( 30.87, 50.14) --
	( 92.06, 50.14);

\path[draw=drawColor,line width= 0.3pt,line join=round] ( 30.87, 62.89) --
	( 92.06, 62.89);

\path[draw=drawColor,line width= 0.3pt,line join=round] ( 30.87, 75.64) --
	( 92.06, 75.64);

\path[draw=drawColor,line width= 0.3pt,line join=round] ( 30.87, 88.39) --
	( 92.06, 88.39);

\path[draw=drawColor,line width= 0.3pt,line join=round] ( 32.94, 28.11) --
	( 32.94, 92.06);

\path[draw=drawColor,line width= 0.3pt,line join=round] ( 47.21, 28.11) --
	( 47.21, 92.06);

\path[draw=drawColor,line width= 0.3pt,line join=round] ( 61.47, 28.11) --
	( 61.47, 92.06);

\path[draw=drawColor,line width= 0.3pt,line join=round] ( 75.73, 28.11) --
	( 75.73, 92.06);

\path[draw=drawColor,line width= 0.3pt,line join=round] ( 90.00, 28.11) --
	( 90.00, 92.06);

\path[draw=drawColor,line width= 0.6pt,line join=round] ( 30.87, 31.01) --
	( 92.06, 31.01);

\path[draw=drawColor,line width= 0.6pt,line join=round] ( 30.87, 43.77) --
	( 92.06, 43.77);

\path[draw=drawColor,line width= 0.6pt,line join=round] ( 30.87, 56.52) --
	( 92.06, 56.52);

\path[draw=drawColor,line width= 0.6pt,line join=round] ( 30.87, 69.27) --
	( 92.06, 69.27);

\path[draw=drawColor,line width= 0.6pt,line join=round] ( 30.87, 82.02) --
	( 92.06, 82.02);

\path[draw=drawColor,line width= 0.6pt,line join=round] ( 40.07, 28.11) --
	( 40.07, 92.06);

\path[draw=drawColor,line width= 0.6pt,line join=round] ( 54.34, 28.11) --
	( 54.34, 92.06);

\path[draw=drawColor,line width= 0.6pt,line join=round] ( 68.60, 28.11) --
	( 68.60, 92.06);

\path[draw=drawColor,line width= 0.6pt,line join=round] ( 82.86, 28.11) --
	( 82.86, 92.06);
\definecolor{fillColor}{RGB}{29,145,192}

\path[fill=fillColor] ( 33.65, 31.01) rectangle ( 46.49, 47.59);

\path[fill=fillColor] ( 47.92, 31.01) rectangle ( 60.76, 89.16);

\path[fill=fillColor] ( 62.18, 31.01) rectangle ( 75.02, 43.26);

\path[fill=fillColor] ( 76.45, 31.01) rectangle ( 89.28, 86.48);
\end{scope}
\begin{scope}
\definecolor{drawColor}{gray}{0.30}

\node[text=drawColor,anchor=base east,inner sep=0pt, outer sep=0pt, scale=  0.80] at ( 25.92, 28.26) {0.0};

\node[text=drawColor,anchor=base east,inner sep=0pt, outer sep=0pt, scale=  0.80] at ( 25.92, 41.01) {0.1};

\node[text=drawColor,anchor=base east,inner sep=0pt, outer sep=0pt, scale=  0.80] at ( 25.92, 53.76) {0.2};

\node[text=drawColor,anchor=base east,inner sep=0pt, outer sep=0pt, scale=  0.80] at ( 25.92, 66.51) {0.3};

\node[text=drawColor,anchor=base east,inner sep=0pt, outer sep=0pt, scale=  0.80] at ( 25.92, 79.26) {0.4};
\end{scope}
\begin{scope}
\definecolor{drawColor}{gray}{0.30}

\node[text=drawColor,anchor=base,inner sep=0pt, outer sep=0pt, scale=  0.80] at ( 40.07, 17.65) {0};

\node[text=drawColor,anchor=base,inner sep=0pt, outer sep=0pt, scale=  0.80] at ( 54.34, 17.65) {1};

\node[text=drawColor,anchor=base,inner sep=0pt, outer sep=0pt, scale=  0.80] at ( 68.60, 17.65) {2};

\node[text=drawColor,anchor=base,inner sep=0pt, outer sep=0pt, scale=  0.80] at ( 82.86, 17.65) {3};
\end{scope}
\begin{scope}
\definecolor{drawColor}{RGB}{0,0,0}

\node[text=drawColor,anchor=base,inner sep=0pt, outer sep=0pt, scale=  0.80] at ( 61.47,  7.44) {Printed Value};
\end{scope}
\begin{scope}
\definecolor{drawColor}{RGB}{0,0,0}

\node[text=drawColor,rotate= 90.00,anchor=base,inner sep=0pt, outer sep=0pt, scale=  0.80] at ( 11.01, 60.09) {Probability};
\end{scope}
\end{tikzpicture}

%% file: 20_autosar.tex
\section{AUTOSAR Adaptive Platform}
\label{sec:autosar}

\subsection{Overview}
\label{sec:autosar-overview}

AUTOSAR Adaptive Platform is a \acf{SoA} that is based on a POSIX-compliant
operating system. The software stack consists of a middleware that handles
communication between services, and the Runtime Environment for Adaptive
Applications (ARA) that provides common APIs and services. While the standard
does not specify the precise middleware and supports third-party solutions, 
the AUTOSAR consortium suggests using the SOME/IP
protocol~\cite{autosar2018-some-protocol-specific,autosar2018-some-service}.

Adaptive AUTOSAR applications are organized in \acfp{SWC} that communicate via
services that they may provide or request.  We call an \ac{SWC} that provides a
service a \emph{server} and an \ac{SWC} that requests a service a
\emph{client}.  Client and server roles may be fulfilled by the same \ac{SWC}.
\acp{SWC} provide or request services as needed; the binding between clients
and servers is determined at runtime by the middleware through service
discovery.  The dynamic binding of services is the core mechanism for providing
adaptivity in \ac{AP}.

The service interfaces are fully specified at design time and are composed of
methods, events, and fields.  While events are one-way messages that the server
initiates and the client handles, methods are two-way messages that the client
initiates and the server responds to. Fields are state variables exposed by the
server. Each field may provide a get method, a set method and an event that
indicates state changes.

\begin{figure}[ht]
  \centering
  \input{fig/autosar_com.tex}
  \vspace{-0.1cm}
  \caption{Communication mechanism in AUTOSAR \ac{AP}.
    Client and server use auto-generated proxies and skeletons to communicate
    with their peers.}
  \label{fig:autosar_com}
  \vspace{-0.4cm}
\end{figure}
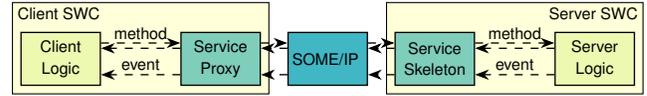

Figure~\ref{fig:autosar_com} illustrates the communication mechanisms of AUTOSAR
\ac{AP}. \acp{SWC} abstract over the precise middleware
by using \emph{proxies} and \emph{skeletons} that are generated from a
service description. A skeleton is an abstract interface that a server needs to
implement in order to provide a service. A proxy is an object that a
client receives when requesting a service.  Client and server communicate directly
through the proxy and skeleton objects. For instance, the invocation of a
method provided by the proxy translates into a message being sent via the
middleware to the server, which then translates the message back into a 
method call. The implementation of the service method is expected to
return a future. As soon as the corresponding promise is fulfilled, the
server sends a message back to the client.

Applications in AUTOSAR \ac{AP} commonly consist of multiple \acp{SWC}.  Each
individual \ac{SWC} can be considered a full program as it is mapped to a
process on the target platform during deployment. While the service-oriented
communication model of AUTOSAR \ac{AP} specifies how \acp{SWC} interact,
it does not specify how \acp{SWC} should be implemented. The standard, however, 
suggests a thread-based coding style.

\subsection{Nondeterminism in AUTOSAR Adaptive Platform}
Despite its goal of supporting safety-critical, possibly autonomous applications,
AUTOSAR \ac{AP} uses a \ac{MoC} that is
inherently nondeterministic. We identify three distinct sources of
nondeterminism in \ac{AP}:
\begin{enumerate}
\item The suggested programming model for the implementation of individual
  \acp{SWC} is based on threads. Threads, however, make it notoriously difficult
  to engineer deterministic concurrent
  software~\cite{lee2006-problem-with-threads}. AUTOSAR \ac{AP} provides coding
  guidelines to avoid the problems with threads, but nondeterminism is still
  likely to creep in, especially when code evolves over
  time~\cite{gu2015-what-change}.
\item The order in which \acp{SWC} process incoming messages
  is undefined. If two clients call the same method on a service, the
  calls may be processed in either order.
\item Point-to-point in-order message delivery could be achieved by the
middleware and underlying TCP/IP network stack, but this is not a formal
requirement in AUTOSAR \ac{AP}. Even if in-order message delivery is guaranteed, the
time required for message transport is still unpredictable.
\end{enumerate}

One provision for deterministic execution that the AUTOSAR \ac{AP} introduces is
the ``deterministic client''~\cite{autosar2019-specific-execution-managemen}, which provides a task-based programming
model for the implementation of \acp{SWC}. Because its scope is limited to
\emph{individual} \acp{SWC}, the solution only addresses the first source of
nondeterminism. Applications that consist of multiple communicating
deterministic clients can still exhibit nondeterminism via 2) and 3). The
solution we outline in the remainder of this paper addresses all three. It
should be noted, however, that the mechanisms for recovering transient errors
and ensuring predictable execution time offered by the deterministic client
can be combined with the solution we propose.

%% file: fig/autosar_com.tex
\begin{tikzpicture}[font=\sffamily]
  \scriptsize

  \tikzstyle{actor}=[draw, rectangle, align=center, minimum height=16pt, minimum width=40pt,line width=0.5pt,fill=c4,inner sep=0]
  \tikzstyle{large}=[actor, minimum height=20pt, minimum width=30pt, inner sep=0pt]
  \tikzstyle{label}=[above=-1pt,midway,color=black]
  \tikzstyle{sync}=[draw,{Triangle[reversed,scale=0.9]}-{Triangle[scale=0.9]},line width=0.5pt]
  \tikzstyle{async}=[draw,dashed,-{Stealth[]},line width=0.5pt]
  \tikzstyle{swc}=[draw, inner sep=3pt, fill=c1]
  \tikzstyle{ethernet}=[draw, minimum width=83pt, minimum height=18pt, fill=c6,rotate=90]

  \node[large,fill=c2] (cl) { Client\\Logic };
  \node[large,right=30pt of cl,fill=c4] (proxy) { Service\\Proxy };
  \node[large,right=10pt of proxy,fill=c5] (some) { SOME/IP };
  \node[large,right=10pt of some,fill=c4] (skel) { Service\\Skeleton };
  \node[large,right=30pt of skel,fill=c2] (sl) { Server\\Logic };

  \path[async] ([yshift=6pt]cl.east) -- node[label]{method} ([yshift=6pt]proxy.west);
  \path[async] ([yshift=4pt]proxy.west) -- ([yshift=4pt]cl.east);
  \path[async] ([yshift=-6pt]proxy.west) -- node[label]{event} ([yshift=-6pt]cl.east);

  \path[async] ([yshift=6pt]proxy.east) -- ([yshift=6pt]some.west);
  \path[async] ([yshift=4pt]some.west) -- ([yshift=4pt]proxy.east);
  \path[async] ([yshift=-6]some.west) -- ([yshift=-6pt]proxy.east);

  \path[async] ([yshift=6pt]some.east) -- ([yshift=6pt]skel.west);
  \path[async] ([yshift=4pt]skel.west) -- ([yshift=4pt]some.east);
  \path[async] ([yshift=-6pt]skel.west) -- ([yshift=-6pt]some.east);

  \path[async] ([yshift=6pt]skel.east) -- node[label]{method} ([yshift=6pt]sl.west);
  \path[async] ([yshift=4pt]sl.west) -- ([yshift=4pt]skel.east);
  \path[async] ([yshift=-6pt]sl.west) -- node[label]{event} ([yshift=-6pt]skel.east);

  \begin{scope}[on background layer]
    \node[above=3pt of cl] (d1) {};
    \node[swc,fit={(cl)(proxy)(d1)}] (cswc) {};
    \node[below right, inner sep=2pt] at (cswc.north west) {Client SWC};
    \node[above=3pt of sl] (d3) {};
    \node[swc,fit={(sl)(skel)(d3)}] (sswc) {};
    \node[below left, inner sep=2pt] at (sswc.north east) {Server SWC};
  \end{scope}

\end{tikzpicture}

%% file: 30_discrete_events.tex
\section{Achieving Determinism in AUTOSAR \ac{AP}}
\label{sec:de}

\subsection{The Reactor Model}

\acp{SWC} in AUTOSAR \ac{AP} follow the design pattern of actors, which are
concurrent stateful processes that communicate via asynchronous message
passing~\cite{hewitt1977viewing,agha1985actors}.  In the actor model, no
constraints are enforced on the ordering of message delivery, but coordination
strategies for achieving deterministic actor programs are
known~\cite{deterministicActors19}.  A recently introduced variation of actors,
called reactors~\cite{lohstroh2019-invited,reactors-cyphy}, comprises an
execution model based on a discrete events semantics that is deterministic by
default, but admits explicit sources of nondeterminism to accommodate sporadic
sensors, interrupts, and other nondeterministic components that are
indispensable in cyber-physical applications like automotive software.

Unlike actors, communications between reactors occur via events that are
associated with tags (also called timestamps). Coordination entails ensuring
that all communication between reactors happens in tag order. Reactors are
composed out of reactions that can be triggered by input events and may produce
output events. The tags of events produced by a reaction are identical to the
tags of the events that triggered it; tags denote \emph{logical} time and
reactions are logically instantaneous. Reactions can also be triggered by
action events, which may emanate from asynchronous resources (e.g., a sporadic
sensor) managed within the reactor.  Such asynchronously scheduled actions,
called \emph{physical actions}, are tagged based on the last observed
\emph{physical} time, potentially with an additional delay.

The tag of a physical action can be considered an input to the reactor system.
Hence, while physical actions may occur unexpectedly, their occurrence does not
compromise the deterministic operation of the system---unless physical actions
are used as end points for a communication channel between reactors in the same
system. Additional coordination is necessary to preserve determinism in that
case.  Precisely because actions can be scheduled sporadically, to ensure
in-order processing of events, no events are handled before physical time
exceeds their tag.

Another difference with actors is that reactors only communicate to one another
via channels that connect reactor ports. The communication topology of a
reactor program translates into an acyclic precedence graph (APG) that drives
the execution. A reactor runtime scheduler is responsible for transparently
exploiting concurrency in the APG by mapping independent reactions to separate
worker threads.  Reactors are capable of performing efficient multi-threaded
computation, while the programmer is freed from the burden of managing the
interactions between threads, effectively addressing the first source of
nondeterminism discussed in Section~\ref{sec:autosar}.

Because events are timed, the reactor model also allows for the expression of
deadlines, making it suitable for specifying programs with real-time
constraints. A deadline \(D\) is considered violated when an event with tag
\(t\) triggers a reaction associated with \(D\) \emph{after} physical time
\(T\) has exceeded \(t+D\).

Importantly, reactors can also be coordinated deterministically across
\acp{SWC}, by leveraging safe-to-process analysis known from
PTIDES~\cite{zhao2007-programm-model,derler2008-ptides}. This addresses the
second and third source of nondeterminism discussed in
Section~\ref{sec:autosar}.  This approach assumes that each network interface
has an associated deadline \(D\). By further assuming that distributed reactors
have synchronized physical clocks with a bounded clock synchronization error
\(E\) (this is the case in \ac{AP}~\cite{autosar2019-specific-time}) and their
communication has a bounded latency \(L\), in-order event handling can be
assured.  Specifically, when a reactor receives a message with tag \(t\) from
the network, it has to schedule an action with tag \(t+D+L+E\) in reaction to
which it can later safely output the message. The physical time delay enforced
by the scheduler ensures that no message with a timestamp smaller than \(t\) is
still expected to arrive over the network.

\begin{figure*}[t]
  \centering
  \input{fig/dear.tex}
  \vspace{-0.2cm}
  \caption{Integration of reactors in AUTOSAR AP. Special reactors (transactors)
    translate between the reactor implementation of the \ac{SWC} logic and the
    service interface that the \ac{SWC} exposes to its environment.}
  \label{fig:dear}
  \vspace{-0.4cm}
\end{figure*}

We propose to use reactors as a programming model for designing
deterministic \acp{SWC} and attach tags to communications between \acp{SWC} to
allow for the preservation of the deterministic discrete event semantics of
reactors across \acp{SWC}.

\subsection{Integrating Reactors with AUTOSAR AP}

We created a framework called DEAR\footnote{Available at:
\url{https://github.com/tud-ccc/dear}}~(Discrete Events for AUTOSAR) that
provides a C++ implementation of the reactor model. It implements type-safe
mechanisms for the definition of reactors with ports, actions and
reactions. This also includes mechanisms for composing reactors to form
deterministic programs.  The framework further provides an implementation of
the runtime scheduler to coordinate the execution of the reactor network.  This
establishes a foundation for the design and execution of deterministic
\acp{SWC}.

In order to enable composition of deterministic applications from deterministic
\acp{SWC}, a mechanism is needed for transporting tagged messages in AUTOSAR
\ac{AP} so that safe-to-process analysis can be performed prior to inserting
events into the receiving reactor network. This is challenging since the
standard for AUTOSAR \ac{AP} explicitly specifies the interface that \acp{SWC}
use for communication. Exposing reactor ports directly to the interface of
\acp{SWC} would break compatibility with the standard.  We can work around this
by introducing transactors that translate between the service-oriented
interfaces of \acp{SWC} and the event-based input and output ports of reactors.

DEAR provides four distinct transactors, each implemented as a reactor and
enabling the composition of reactors through regular AUTOSAR service
interfaces.  The \emph{client method transactor} interacts with a given method
of a service interface in the client role. Similarly, the \emph{server method
transactor} interacts with a method in the server role. Analogous to methods, a
similar pair of transactors for interacting with \ac{AP} events in the role of
clients and servers exists. Since fields are composed of a get method, a set
method and an event, interaction with fields requires the use of one event and
two method transactors. All four transactors are shown in
Figure~\ref{fig:dear}.  Given a service interface, the transactors required for
interacting via this particular interface can be automatically generated.

AUTOSAR \ac{AP} provides no mechanisms for associating metadata like reactor
tags with method calls or events. This, however, is required for the
transmission of tagged messages between deterministic \acp{SWC}. Therefore, we
modified the library that binds to the SOME/IP middleware to optionally append
tags to outgoing messages and to retrieve tags from incoming messages if
available. This modification is not in violation of the standard. It can be
seen as the introduction of a new third-party middleware that extends over
SOME/IP by allowing the transmission of tagged messages. While the transactors
use the regular AUTOSAR AP service proxies and skeletons, for each event
occurrence or method call they store a corresponding tag that is picked by the
modified SOME/IP middleware prior to transmitting the payload over the network.

The entire process of transmitting tagged messages between \acp{SWC} in DEAR is
illustrated in Figure~\ref{fig:dear}.  The sequence starts with a client that
invokes a method call on a remote service.  In the reactor implementation, that
corresponds to producing an event with tag \(t_c\) on the output port connected
to the request input port of the client method transactor~(1). This input port
has a configurable deadline \(D_c\). If the deadline is not violated, the
corresponding reaction sends \(t_c+D_c\) to the timestamp bypass~(2) and
invokes the actual method call on the service proxy object~(3). Thereby, it
forwards the data associated with the incoming event as method arguments. The
service proxy calls the SOME/IP binding~(4) to prepare a network message. Our
modified binding retrieves \(t_c+D_c\) from the timestamp bypass~(5) and
attaches it to the SOME/IP message, which it then sends over the network to the
server~(6).

Upon receiving the network message on the server side, our modified SOME/IP
binding retrieves \(t_c+D_c\) from the message and sends it to the local
timestamp bypass~(7) before invoking the corresponding method call on the
service skeleton~(8). This method call triggers an interrupt of the server
method transactor~(9) which retrieves \(t_c+D_c\) from the timestamp
bypass~(10) and schedules an action with tag \(t_c+D_c+L+E\) accounting for the
worst-case network latency \(L\), as well as the maximum clock skew \(E\)
between platforms.  The reaction to this action produces an event on the output
port that forwards the method arguments to the reactor that implements the
server logic~(11). The server logic then reacts to this event.

The server eventually sends a response by producing an event with tag \(t_s\)
(\(t_s \ge t_c+D_c+L+E\)) on the output port connected to the input port of the
service method transactor~(12). This input port has a configurable deadline
\(D_s\). If the deadline is not violated, the corresponding reaction sends
\(t_s+D_s\) to the timestamp bypass~(13) and returns the data associated with
the event to the service skeleton~(14). The service skeleton calls the SOME/IP
binding~(15) to create a response message. The binding retrieves \(t_s+D_s\)
from the timestamp bypass~(16) and attaches it to the outgoing message which it
then sends over the network to the client~(17).

The client SOME/IP binding retrieves \(t_s+D_s\) from the message and sends it
to the timestamp bypass~(18) while forwarding the return value to the service
proxy (19). The arrival of the return value triggers an interrupt in the client
method transactor~(20) which retrieves \(t_s+D_s\) from the timestamp bypass
(21) and schedules an action with tag \(t_s+D_s+L+E\) to again account for
transmission latency and clock synchronization error. Finally, the reaction for
this action produces an event on the output port of the transactor~(22).

While transparently enabling deterministic composition of reactor-based
\acp{SWC}, our solution leaves open the possibility of composing reactor-based
\acp{SWC} with regular service implementations that do not communicate via
tagged signals.  The default behavior of our transactors is to fail when
receiving messages without an associated timestamp, but they can also be
configured to tag received messages with the physical time at which they are
received. This approach treats the arrival of untagged messages the same way as
reactors deal with the arrival of sporadic sensor readings. This essentially
furnishes backward compatibility with existing service implementations and the
ability to gradually introduce reactor-based \acp{SWC}.

%% file: fig/dear.tex
\begin{tikzpicture}[font=\sffamily]
  \scriptsize

  \tikzstyle{actor}=[draw, rectangle, align=center, minimum height=16pt, minimum width=44pt,line width=0.5pt,fill=c4,inner sep=0]
  \tikzstyle{large}=[actor, minimum height=37pt, minimum width=30pt]
  \tikzstyle{sync}=[draw,{Triangle[reversed,scale=0.9]}-{Triangle[scale=0.9]},line width=0.5pt]
  \tikzstyle{label}=[above=-1pt,midway,color=black,font=\tiny]
  \tikzstyle{blabel}=[below=-1pt,midway,color=black,font=\tiny]
  \tikzstyle{async}=[draw,dashed,-{Stealth[]},line width=0.5pt]
  \tikzstyle{swc}=[draw, inner sep=2pt, fill=c1]
  \tikzstyle{ethernet}=[draw, minimum width=63pt, minimum height=18pt, fill=c6,rotate=90]

  \node[minimum height=0,inner sep=0] (canchor) {};

  \node[actor,above=2pt of canchor, fill=c3] (cmt) {Client Method\\Transactor};
  \node[actor,below=2pt of canchor,fill=c3] (cet) {Client Event\\Transactor};
  \node[large,left=40pt of canchor,fill=c2] (cl) { Client\\Reactor };
  \path[sync] ([yshift=12.5pt]cl.east) -- node [label] {(1)} ([yshift=2pt]cmt.west);
  \path[sync] ([yshift=-2pt]cmt.west) -- node [blabel] {(22)} ([yshift=8.5pt]cl.east);
  \path[sync] (cet.west) -- ([yshift=-10.5pt]cl.east);

  \node[large,right=40pt of canchor,fill=c4] (proxy) { Service\\Proxy };
  \path[async] ([yshift=1pt]cmt.east) -- node[label]{(3)} ([yshift=11.5pt]proxy.west);
  \path[async] ([yshift=9.5pt]proxy.west) -- node[blabel]{(20)} ([yshift=-1pt]cmt.east);
  \path[async] ([yshift=-10.5pt]proxy.west) -- (cet.east);

  \node[large,right=20pt of proxy,fill=c5] (cbind) { SOME/IP\\Binding };
  \path[async] ([yshift=11.5pt]proxy.east) -- node[label]{(4)} ([yshift=11.5pt]cbind.west);
  \path[async] ([yshift=9.5pt]cbind.west) -- node[blabel]{(19)} ([yshift=9.5pt]proxy.east);
  \path[async] ([yshift=-10.5pt]cbind.west) -- ([yshift=-10.5pt]proxy.east);

  \node[actor,above=4pt of proxy] (ctb) {Timestamp\\Bypass};
  \path[async] let \p1=([xshift=-1pt]cmt.north), \p2=([yshift=1pt]ctb.west) in (\p1) -- (\x1,\y2)-- node[label]{(2)} (\p2);
  \path[async] let \p1=([xshift=1pt]cmt.north), \p2=([yshift=-1pt]ctb.west) in (\p2) --  node[blabel]{(21)} (\x1,\y2)-- (\p1);
  \path[async] let \p1=([yshift=1pt]ctb.east), \p2=([xshift=1pt]cbind.north) in (\p1) -- node[label]{(5)} (\x2,\y1)-- (\p2);
  \path[async] let \p1=([yshift=-1pt]ctb.east), \p2=([xshift=-1pt]cbind.north) in (\p2) -- (\x2,\y1)-- node[blabel]{(18)} (\p1);
  \path[async,-] ([xshift=1pt]cmt.south) -- ([xshift=1pt]cet.north);
  \path[async,-] ([xshift=-1pt]cet.north) -- ([xshift=-1pt]cmt.south);

  \begin{scope}[on background layer]
    \node[swc,fit={(cl)(proxy)(cbind)(ctb)}] (cswc) {};
    \node[below right, inner sep=4pt] at (cswc.north west) {Client SWC};
  \end{scope}

  \node[minimum height=0,inner sep=0,right=300pt of canchor] (sanchor) {};

  \node[actor, above=2pt of sanchor,fill=c3] (cmt) {Server Method\\Transactor};
  \node[actor,below=2pt of sanchor,fill=c3] (scet) {Server Event\\Transactor};
  \node[large,right=40pt of sanchor,fill=c2] (sl) { Server\\Reactor };
  \path[sync] ([yshift=2pt]cmt.east) -- node [label] {(11)} ([yshift=12.5pt]sl.west);
  \path[sync] ([yshift=8.5pt]sl.west) -- node [blabel] {(12)} ([yshift=-2pt]cmt.east);
  \path[sync] ([yshift=-10.5pt]sl.west) -- (scet.east);

  \node[large,left=40pt of sanchor,fill=c4] (skel) { Service\\Skeleton };
  \path[async] ([yshift=11.5pt]skel.east) -- node[label]{(9)} ([yshift=1pt]cmt.west);
  \path[async] ([yshift=-1pt]cmt.west) -- node[blabel]{(14)}  ([yshift=9.5pt]skel.east);
  \path[async] (scet.west) -- ([yshift=-10.5pt]skel.east);

  \node[large,left=20pt of skel,fill=c5] (sbind) { SOME/IP\\Binding };
  \path[async] ([yshift=11.5pt]sbind.east) -- node[label]{(8)} ([yshift=11.5pt]skel.west);
  \path[async] ([yshift=9.5pt]skel.west) -- node[blabel]{(15)} ([yshift=9.5pt]sbind.east);
  \path[async] ([yshift=-10.5pt]skel.west) -- ([yshift=-10.5pt]sbind.east);

  \node[actor,above=4pt of skel] (stb) {Timestamp\\Bypass};
  \path[async] let \p1=([xshift=-1pt]cmt.north), \p2=([yshift=-1pt]stb.east) in (\p1) -- (\x1,\y2)-- node[blabel]{(13)} (\p2);
  \path[async] let \p1=([xshift=1pt]cmt.north), \p2=([yshift=1pt]stb.east) in (\p2) --  node[label]{(10)} (\x1,\y2)-- (\p1);
  \path[async] let \p1=([yshift=-1pt]stb.west), \p2=([xshift=1pt]sbind.north) in (\p1) -- node[blabel]{(16)} (\x2,\y1)-- (\p2);
  \path[async] let \p1=([yshift=1pt]stb.west), \p2=([xshift=-1pt]sbind.north) in (\p2) -- (\x2,\y1)-- node[label]{(7)} (\p1);
  \path[async,-] ([xshift=1pt]cmt.south) -- ([xshift=1pt]scet.north);
  \path[async,-] ([xshift=-1pt]scet.north) -- ([xshift=-1pt]cmt.south);

  \begin{scope}[on background layer]
    \node[swc,fit={(sl)(skel)(sbind)(stb)}] (sswc) {};
    \node[below left, inner sep=4pt] at (sswc.north east) {Server SWC};
  \end{scope}

  \node[ethernet] (eth) at ($(cswc)!0.5!(sswc)$) {Ethernet};
  \begin{scope}[on background layer]
  \path[async] ([yshift=11.5pt]cbind.east) -- node[label]{(6)} ([yshift=1.5pt]eth.north);
  \path[async] ([yshift=-0.5pt]eth.north) -- node[blabel]{(17)}  ([yshift=9.5pt]cbind.east);
  \path[async] ([yshift=-20.5pt]eth.north) -- ([yshift=-10.5pt]cbind.east);
  \path[async] ([yshift=1.5pt]eth.south) -- node[label]{(6)} ([yshift=11.5pt]sbind.west);
  \path[async] ([yshift=9.5pt]sbind.west) -- node[blabel]{(17)} ([yshift=-0.5pt]eth.south);
  \path[async] ([yshift=-10.5pt]sbind.west) -- ([yshift=-20.5pt]eth.south);
  \end{scope}
\end{tikzpicture}

%% file: 40_case_study.tex
\section{Case Study: Adaptive Platform Demonstrator}
\label{sec:case-study}

\subsection{Nondeterministic Brake Assistant}

The AUTOSAR consortium provides the \acf{APD}, which is an example
implementation of the specification for AUTOSAR \ac{AP}.  It provides a set of
demo applications, where the most realistic and advanced application is the
brake assistant shown in Figure~\ref{fig:eba}. Unlike what may be expected of a
safety-critical application, the brake assistant exhibits nondeterminism that
could potentially have fatal consequences.  While this demo is not designed for
deployment in the real world, it illuminates the presence of uncontrollable and
safety-undermining nondeterminism in \ac{AP}.

\begin{figure}[ht]
  \vspace{-0.1cm}
  \centering
    \input{fig/eba.tex}
  \vspace{-0.1cm}
    \caption{Brake assistant application in APD}
  \label{fig:eba}
\end{figure}

The brake assistant consists of a pipeline of five \acp{SWC}, distributed
across two platforms. {\sf Video Provider} captures video frames and sends one
approximately every \SI{50}{ms} (via a proprietary protocol) to {\sf Video
Adapter}, which is running on the second platform. The communication along the
remainder of the component chain occurs through \ac{AP} service interfaces via
the SOME/IP middleware. Event notifications are used to transfer data from one
\ac{SWC} to the next and the corresponding event handler stores the data in a
one-slot input buffer. Each \ac{SWC} sets up a periodic callback so that the OS
triggers the \ac{SWC} logic every \SI{50}{ms}. Each component then reads the
current data item from its input buffer, performs some computations, and then
communicates the result via an event.

For each frame {\sf Preprocessing} receives from {\sf Video Adapter}, it
computes a bounding box demarcating the current travel lane. {\sf Computer
Vision} receives from {\sf Preprocessing} both the lane information as well as
the original frame and uses it to detect vehicles in the lane and estimates
their distance. It forwards the list of detected vehicles to the {\sf EBA}
component, which in turn decides whether an emergency brake maneuver is
required.

The logic of each component processes the last data written to its one-slot
input buffer. If there is no data, the \acp{SWC} silently stop computation and
wait for the next periodic trigger to occur.  This introduces nondeterminism as
data could get overwritten before it is read by a downstream component, causing
entire frames to be dropped. Moreover, since the Computer Vision component
reads not one but two inputs, this can lead to misalignment between the video
frames and the lane information. We instrumented the brake assistant code to
detect and report frame loss and misalignment.  Execution on our evaluation
platform, consisting of two MinnowBoard
Turbot\footnote{\url{https://minnowboard.org/minnowboard-turbot-dual-e/technical-specs}}
boards connected via an Ethernet switch, confirmed that the described errors
indeed occur in a real-world setting. The boards are equipped with an Intel
Atom E3845 Quad-Core processor and are officially supported by the \ac{APD}.

\begin{figure}[t]
  \centering
  \input{plot/errors.tex}
  \vspace{-0.2cm}
  \caption{Prevalence of errors for 20 executions of the brake assistant.}
  \label{fig:errors}
  \vspace{-0.4cm}
\end{figure}
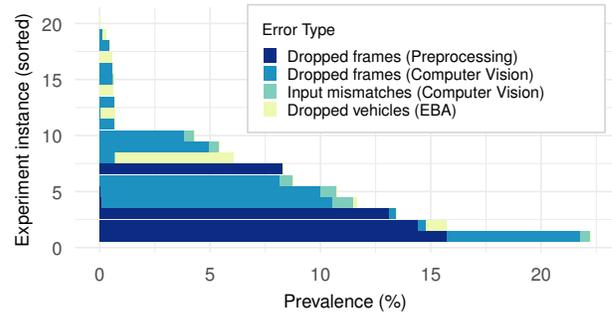

We performed a series of experiments to analyze the prevalence of the described
errors. We let the brake assistant processes a total of 100,000 frames and
counted dropped inputs and mismatches.  Figure~\ref{fig:errors} plots the
obtained results for a total of 20 experiments. Each bar in the figure shows
the error prevalence for one instance of the experiment. The results are
ordered by error rate for better visibility.

The error rate varies significantly between experiments. In the best case we
observed an error rate of \(0.018\%\) and in the worst case an error rate of
\(22.25\%\). On average we observed \(5.60\%\) errors. Also the composition of
error types varies significantly. In most experiments frame dropping at {\sf
Computer Vision} was dominant, but sometimes dropped vehicles at {\sf EBA} or
dropped frames at {\sf Preprocessing} dominated.  This underlines the
difficulty of assessing the performance and correctness of the brake
assistant. It appears the error rate is strongly influenced by the offset
between the individual periodic callbacks of the \acp{SWC}, which depends on
when \acp{SWC} are started and is difficult to control.

\subsection{Deterministic Brake Assistant}

DEAR allows us to easily transform the brake assistant application into a
deterministic reactor implementation. Since the original implementation
separates computational logic from the communication mechanism, transformation
requires only a few code changes.  We encapsulate the logic of each \ac{SWC} in
a reactor that has one reaction to process incoming events. This reaction calls
the original logic to process the data associated with the incoming event and
produces an output event.  In order to support the transmission of tagged
messages between \acp{SWC}, each reactor binds to the service interfaces of the
\ac{SWC} using the DEAR transactors. As described in Section~\ref{sec:de}, a
carefully chosen deadline on the reaction of each sending transactor ensures
that there is an upper bound on how much logical time lags behind physical
time. The receiving transactor further accounts for the physical delay of
message transmission and ensures that incoming messages are only processed when
it is safe to do so.

Since {\sf Computer Vision} has two inputs, the reaction that calls its logic
expects to receive two events with the same tag at both inputs.  If only one
input is received, this is considered an error. {\sf Video Adapter} has no
well-defined input. It sporadically receives frames over the network sent by
the camera.  As the timing of the camera cannot be controlled, we implement
{\sf Video Adapter} as a sensor that inserts frames into the reactor network
with a tag equal to the physical time of message reception.  Once the incoming
frame is tagged, subsequent reactions are carried out in a deterministic order.

In order to achieve correct execution, it is important to carefully consider
the physical delays imposed by the computations of each \ac{SWC} as well as by
the transport of messages between \acp{SWC}. Only if the deadlines associated
with each \ac{SWC} account for its WCET and the specified maximum communication
latency and synchronization error are accurate, correct execution is
guaranteed. In our implementation, we set the deadlines to \SI{5}{ms} for {\sf
Video Adapter}, \SI{25}{ms} for {\sf Preprocessing}, \SI{25}{ms} for {\sf
Computer Vision} and \SI{5}{ms} for {\sf EBA}. We further assume a maximum
communication latency of \SI{5}{ms}. Since all \acp{SWC} of this application
are deployed to the same platform, there is no clock synchronization error to
account for. Note that these numbers are estimated upper bounds of delays. More
precise values can be obtained from WCET analysis.

With this implementation, we achieve correct and deterministic execution on the
MinnowBoard platform. Moreover, the timed semantics of reactors facilitates
reasoning about the worst-case end-to-end latency between receiving a frame and
producing an output brake signal.  These benefits come at the cost of an extra
physical time delay as each \ac{SWC} needs to account for worst case
computation and communication delays.  This, however, is not a necessity. For
certain applications it is acceptable to deliberately introduce the possibility
of sporadic errors by setting deadlines to values lower than the actual WCET.
Independent from how deadlines and communication delay are chosen, the reactor
semantics guarantees determinism and translates any violation of one of the
assumptions directly into observable errors. In contrast to the original brake
assistant implementation, the trade-off between end-to-end latency and error
rate becomes apparent.

%% file: fig/eba.tex
\begin{tikzpicture}[font=\sffamily]
  \scriptsize

  \tikzstyle{platform}=[draw, inner sep=2pt, fill=c2]
  \tikzstyle{swc}=[draw, fill=c5,align=center,minimum width=38pt, minimum height=16pt, inner sep=0]

  \node[swc] (vp) { Video\\Provider };

  \node[swc,right=9pt of vp] (va) { Video\\Adapter };
  \node[swc,right=3pt of va] (pp) { Pre-\\processing };
  \node[swc,right=3pt of pp] (cv) { Computer \\ Vision };
  \node[swc,right=3pt of cv] (eba) { EBA };

  \begin{scope}[on background layer]
  \node[above=5pt of vp] (d1) {};
  \node[platform, fit=(vp)(d1)] (p1) {};
  \node[below left, inner sep=2pt] at (p1.north east) {Platform 1};
  \node[above=5pt of eba] (d2) {};
  \node[platform, align=left,fit={(va)(pp)(cv)(eba)(d2)}] (p2) {};
  \node[below left, inner sep=2pt] at (p2.north east) {Platform 2};
  \end{scope}

  \path[draw,-latex] ([xshift=10pt]vp.south) to[bend right=45] node[below,midway]{frame} ([xshift=-10pt]va.south);
  \path[draw,-latex] ([xshift=10pt]va.south) to[bend right=45] node[below,midway]{frame} ([xshift=-10pt]pp.south);
  \path[draw,-latex] ([xshift=10pt]pp.south) to[bend right=45] node[below,midway]{frame} ([xshift=-10pt]cv.south);
  \path[draw,-latex] ([xshift=10pt]pp.north) to[bend left=45] node[above,midway]{lane} ([xshift=-10pt]cv.north);
  \path[draw,-latex] ([xshift=10pt]cv.south) to[bend right=45] node[below,midway]{vehicles} ([xshift=-10pt]eba.south);
  \path[draw,-latex] ([xshift=10pt]eba.south) to[bend right=45] node[below,midway]{brake} ([xshift=29pt]eba.south);

\end{tikzpicture}

%% file: plot/errors.tex
\begin{tikzpicture}[x=1pt,y=1pt,font=\sffamily]
\definecolor{fillColor}{RGB}{255,255,255}
\begin{scope}
\definecolor{drawColor}{gray}{0.92}

\path[draw=drawColor,line width= 0.3pt,line join=round] ( 28.65, 40.60) --
	(232.99, 40.60);

\path[draw=drawColor,line width= 0.3pt,line join=round] ( 28.65, 61.81) --
	(232.99, 61.81);

\path[draw=drawColor,line width= 0.3pt,line join=round] ( 28.65, 83.02) --
	(232.99, 83.02);

\path[draw=drawColor,line width= 0.3pt,line join=round] ( 28.65,104.24) --
	(232.99,104.24);

\path[draw=drawColor,line width= 0.3pt,line join=round] ( 58.81, 28.11) --
	( 58.81,120.97);

\path[draw=drawColor,line width= 0.3pt,line join=round] (100.55, 28.11) --
	(100.55,120.97);

\path[draw=drawColor,line width= 0.3pt,line join=round] (142.29, 28.11) --
	(142.29,120.97);

\path[draw=drawColor,line width= 0.3pt,line join=round] (184.03, 28.11) --
	(184.03,120.97);

\path[draw=drawColor,line width= 0.3pt,line join=round] (225.76, 28.11) --
	(225.76,120.97);

\path[draw=drawColor,line width= 0.6pt,line join=round] ( 28.65, 30.00) --
	(232.99, 30.00);

\path[draw=drawColor,line width= 0.6pt,line join=round] ( 28.65, 51.21) --
	(232.99, 51.21);

\path[draw=drawColor,line width= 0.6pt,line join=round] ( 28.65, 72.42) --
	(232.99, 72.42);

\path[draw=drawColor,line width= 0.6pt,line join=round] ( 28.65, 93.63) --
	(232.99, 93.63);

\path[draw=drawColor,line width= 0.6pt,line join=round] ( 28.65,114.84) --
	(232.99,114.84);

\path[draw=drawColor,line width= 0.6pt,line join=round] ( 37.94, 28.11) --
	( 37.94,120.97);

\path[draw=drawColor,line width= 0.6pt,line join=round] ( 79.68, 28.11) --
	( 79.68,120.97);

\path[draw=drawColor,line width= 0.6pt,line join=round] (121.42, 28.11) --
	(121.42,120.97);

\path[draw=drawColor,line width= 0.6pt,line join=round] (163.16, 28.11) --
	(163.16,120.97);

\path[draw=drawColor,line width= 0.6pt,line join=round] (204.90, 28.11) --
	(204.90,120.97);
\definecolor{fillColor}{RGB}{12,44,132}

\path[fill=fillColor] ( 37.94, 32.33) rectangle (169.45, 36.15);
\definecolor{fillColor}{RGB}{29,145,192}

\path[fill=fillColor] (169.45, 32.33) rectangle (219.60, 36.15);
\definecolor{fillColor}{RGB}{127,205,187}

\path[fill=fillColor] (219.60, 32.33) rectangle (223.69, 36.15);
\definecolor{fillColor}{RGB}{237,248,177}

\path[fill=fillColor] (223.69, 32.33) rectangle (223.70, 36.15);
\definecolor{fillColor}{RGB}{12,44,132}

\path[fill=fillColor] ( 37.94, 36.57) rectangle (158.47, 40.39);
\definecolor{fillColor}{RGB}{29,145,192}

\path[fill=fillColor] (158.47, 36.57) rectangle (161.31, 40.39);
\definecolor{fillColor}{RGB}{127,205,187}

\path[fill=fillColor] (161.31, 36.57) rectangle (161.35, 40.39);
\definecolor{fillColor}{RGB}{237,248,177}

\path[fill=fillColor] (161.35, 36.57) rectangle (169.13, 40.39);
\definecolor{fillColor}{RGB}{12,44,132}

\path[fill=fillColor] ( 37.94, 40.81) rectangle (147.50, 44.63);
\definecolor{fillColor}{RGB}{29,145,192}

\path[fill=fillColor] (147.50, 40.81) rectangle (150.08, 44.63);
\definecolor{fillColor}{RGB}{127,205,187}

\path[fill=fillColor] (150.08, 40.81) rectangle (150.12, 44.63);
\definecolor{fillColor}{RGB}{237,248,177}

\path[fill=fillColor] (150.12, 40.81) rectangle (150.13, 44.63);
\definecolor{fillColor}{RGB}{12,44,132}

\path[fill=fillColor] ( 37.94, 45.06) rectangle ( 38.46, 48.87);
\definecolor{fillColor}{RGB}{29,145,192}

\path[fill=fillColor] ( 38.46, 45.06) rectangle (126.12, 48.87);
\definecolor{fillColor}{RGB}{127,205,187}

\path[fill=fillColor] (126.12, 45.06) rectangle (133.70, 48.87);
\definecolor{fillColor}{RGB}{237,248,177}

\path[fill=fillColor] (133.70, 45.06) rectangle (135.12, 48.87);
\definecolor{fillColor}{RGB}{12,44,132}

\path[fill=fillColor] ( 37.94, 49.30) rectangle ( 38.27, 53.12);
\definecolor{fillColor}{RGB}{29,145,192}

\path[fill=fillColor] ( 38.27, 49.30) rectangle (121.58, 53.12);
\definecolor{fillColor}{RGB}{127,205,187}

\path[fill=fillColor] (121.58, 49.30) rectangle (127.67, 53.12);
\definecolor{fillColor}{RGB}{237,248,177}

\path[fill=fillColor] (127.67, 49.30) rectangle (127.67, 53.12);
\definecolor{fillColor}{RGB}{12,44,132}

\path[fill=fillColor] ( 37.94, 53.54) rectangle ( 37.95, 57.36);
\definecolor{fillColor}{RGB}{29,145,192}

\path[fill=fillColor] ( 37.95, 53.54) rectangle (105.90, 57.36);
\definecolor{fillColor}{RGB}{127,205,187}

\path[fill=fillColor] (105.90, 53.54) rectangle (111.11, 57.36);
\definecolor{fillColor}{RGB}{237,248,177}

\path[fill=fillColor] (111.11, 53.54) rectangle (111.13, 57.36);
\definecolor{fillColor}{RGB}{12,44,132}

\path[fill=fillColor] ( 37.94, 57.78) rectangle (107.15, 61.60);
\definecolor{fillColor}{RGB}{29,145,192}

\path[fill=fillColor] (107.15, 57.78) rectangle (107.20, 61.60);
\definecolor{fillColor}{RGB}{127,205,187}

\path[fill=fillColor] (107.20, 57.78) rectangle (107.20, 61.60);
\definecolor{fillColor}{RGB}{237,248,177}

\path[fill=fillColor] (107.20, 57.78) rectangle (107.20, 61.60);
\definecolor{fillColor}{RGB}{12,44,132}

\path[fill=fillColor] ( 37.94, 62.03) rectangle ( 37.95, 65.84);
\definecolor{fillColor}{RGB}{29,145,192}

\path[fill=fillColor] ( 37.95, 62.03) rectangle ( 43.67, 65.84);
\definecolor{fillColor}{RGB}{127,205,187}

\path[fill=fillColor] ( 43.67, 62.03) rectangle ( 43.75, 65.84);
\definecolor{fillColor}{RGB}{237,248,177}

\path[fill=fillColor] ( 43.75, 62.03) rectangle ( 88.54, 65.84);
\definecolor{fillColor}{RGB}{12,44,132}

\path[fill=fillColor] ( 37.94, 66.27) rectangle ( 37.96, 70.09);
\definecolor{fillColor}{RGB}{29,145,192}

\path[fill=fillColor] ( 37.96, 66.27) rectangle ( 79.21, 70.09);
\definecolor{fillColor}{RGB}{127,205,187}

\path[fill=fillColor] ( 79.21, 66.27) rectangle ( 83.18, 70.09);
\definecolor{fillColor}{RGB}{237,248,177}

\path[fill=fillColor] ( 83.18, 66.27) rectangle ( 83.18, 70.09);
\definecolor{fillColor}{RGB}{12,44,132}

\path[fill=fillColor] ( 37.94, 70.51) rectangle ( 37.95, 74.33);
\definecolor{fillColor}{RGB}{29,145,192}

\path[fill=fillColor] ( 37.95, 70.51) rectangle ( 69.79, 74.33);
\definecolor{fillColor}{RGB}{127,205,187}

\path[fill=fillColor] ( 69.79, 70.51) rectangle ( 73.75, 74.33);
\definecolor{fillColor}{RGB}{237,248,177}

\path[fill=fillColor] ( 73.75, 70.51) rectangle ( 73.75, 74.33);
\definecolor{fillColor}{RGB}{12,44,132}

\path[fill=fillColor] ( 37.94, 74.75) rectangle ( 37.95, 78.57);
\definecolor{fillColor}{RGB}{29,145,192}

\path[fill=fillColor] ( 37.95, 74.75) rectangle ( 43.72, 78.57);
\definecolor{fillColor}{RGB}{127,205,187}

\path[fill=fillColor] ( 43.72, 74.75) rectangle ( 43.78, 78.57);
\definecolor{fillColor}{RGB}{237,248,177}

\path[fill=fillColor] ( 43.78, 74.75) rectangle ( 43.78, 78.57);
\definecolor{fillColor}{RGB}{12,44,132}

\path[fill=fillColor] ( 37.94, 78.99) rectangle ( 38.07, 82.81);
\definecolor{fillColor}{RGB}{29,145,192}

\path[fill=fillColor] ( 38.07, 78.99) rectangle ( 38.21, 82.81);
\definecolor{fillColor}{RGB}{127,205,187}

\path[fill=fillColor] ( 38.21, 78.99) rectangle ( 38.22, 82.81);
\definecolor{fillColor}{RGB}{237,248,177}

\path[fill=fillColor] ( 38.22, 78.99) rectangle ( 43.62, 82.81);
\definecolor{fillColor}{RGB}{12,44,132}

\path[fill=fillColor] ( 37.94, 83.24) rectangle ( 37.95, 87.05);
\definecolor{fillColor}{RGB}{29,145,192}

\path[fill=fillColor] ( 37.95, 83.24) rectangle ( 43.50, 87.05);
\definecolor{fillColor}{RGB}{127,205,187}

\path[fill=fillColor] ( 43.50, 83.24) rectangle ( 43.57, 87.05);
\definecolor{fillColor}{RGB}{237,248,177}

\path[fill=fillColor] ( 43.57, 83.24) rectangle ( 43.57, 87.05);
\definecolor{fillColor}{RGB}{12,44,132}

\path[fill=fillColor] ( 37.94, 87.48) rectangle ( 37.96, 91.30);
\definecolor{fillColor}{RGB}{29,145,192}

\path[fill=fillColor] ( 37.96, 87.48) rectangle ( 38.01, 91.30);
\definecolor{fillColor}{RGB}{127,205,187}

\path[fill=fillColor] ( 38.01, 87.48) rectangle ( 38.01, 91.30);
\definecolor{fillColor}{RGB}{237,248,177}

\path[fill=fillColor] ( 38.01, 87.48) rectangle ( 43.11, 91.30);
\definecolor{fillColor}{RGB}{12,44,132}

\path[fill=fillColor] ( 37.94, 91.72) rectangle ( 37.95, 95.54);
\definecolor{fillColor}{RGB}{29,145,192}

\path[fill=fillColor] ( 37.95, 91.72) rectangle ( 43.03, 95.54);
\definecolor{fillColor}{RGB}{127,205,187}

\path[fill=fillColor] ( 43.03, 91.72) rectangle ( 43.09, 95.54);
\definecolor{fillColor}{RGB}{237,248,177}

\path[fill=fillColor] ( 43.09, 91.72) rectangle ( 43.11, 95.54);
\definecolor{fillColor}{RGB}{12,44,132}

\path[fill=fillColor] ( 37.94, 95.96) rectangle ( 37.95, 99.78);
\definecolor{fillColor}{RGB}{29,145,192}

\path[fill=fillColor] ( 37.95, 95.96) rectangle ( 43.01, 99.78);
\definecolor{fillColor}{RGB}{127,205,187}

\path[fill=fillColor] ( 43.01, 95.96) rectangle ( 43.06, 99.78);
\definecolor{fillColor}{RGB}{237,248,177}

\path[fill=fillColor] ( 43.06, 95.96) rectangle ( 43.07, 99.78);
\definecolor{fillColor}{RGB}{12,44,132}

\path[fill=fillColor] ( 37.94,100.21) rectangle ( 37.96,104.02);
\definecolor{fillColor}{RGB}{29,145,192}

\path[fill=fillColor] ( 37.96,100.21) rectangle ( 37.99,104.02);
\definecolor{fillColor}{RGB}{127,205,187}

\path[fill=fillColor] ( 37.99,100.21) rectangle ( 37.99,104.02);
\definecolor{fillColor}{RGB}{237,248,177}

\path[fill=fillColor] ( 37.99,100.21) rectangle ( 42.77,104.02);
\definecolor{fillColor}{RGB}{12,44,132}

\path[fill=fillColor] ( 37.94,104.45) rectangle ( 37.95,108.27);
\definecolor{fillColor}{RGB}{29,145,192}

\path[fill=fillColor] ( 37.95,104.45) rectangle ( 41.82,108.27);
\definecolor{fillColor}{RGB}{127,205,187}

\path[fill=fillColor] ( 41.82,104.45) rectangle ( 41.86,108.27);
\definecolor{fillColor}{RGB}{237,248,177}

\path[fill=fillColor] ( 41.86,104.45) rectangle ( 41.88,108.27);
\definecolor{fillColor}{RGB}{12,44,132}

\path[fill=fillColor] ( 37.94,108.69) rectangle ( 38.02,112.51);
\definecolor{fillColor}{RGB}{29,145,192}

\path[fill=fillColor] ( 38.02,108.69) rectangle ( 38.95,112.51);
\definecolor{fillColor}{RGB}{127,205,187}

\path[fill=fillColor] ( 38.95,108.69) rectangle ( 39.03,112.51);
\definecolor{fillColor}{RGB}{237,248,177}

\path[fill=fillColor] ( 39.03,108.69) rectangle ( 40.39,112.51);
\definecolor{fillColor}{RGB}{12,44,132}

\path[fill=fillColor] ( 37.94,112.93) rectangle ( 37.96,116.75);
\definecolor{fillColor}{RGB}{29,145,192}

\path[fill=fillColor] ( 37.96,112.93) rectangle ( 38.07,116.75);
\definecolor{fillColor}{RGB}{127,205,187}

\path[fill=fillColor] ( 38.07,112.93) rectangle ( 38.07,116.75);
\definecolor{fillColor}{RGB}{237,248,177}

\path[fill=fillColor] ( 38.07,112.93) rectangle ( 38.08,116.75);
\end{scope}
\begin{scope}
\definecolor{drawColor}{gray}{0.30}

\node[text=drawColor,anchor=base east,inner sep=0pt, outer sep=0pt, scale=  0.80] at ( 23.70, 27.24) {0};

\node[text=drawColor,anchor=base east,inner sep=0pt, outer sep=0pt, scale=  0.80] at ( 23.70, 48.45) {5};

\node[text=drawColor,anchor=base east,inner sep=0pt, outer sep=0pt, scale=  0.80] at ( 23.70, 69.66) {10};

\node[text=drawColor,anchor=base east,inner sep=0pt, outer sep=0pt, scale=  0.80] at ( 23.70, 90.88) {15};

\node[text=drawColor,anchor=base east,inner sep=0pt, outer sep=0pt, scale=  0.80] at ( 23.70,112.09) {20};
\end{scope}
\begin{scope}
\definecolor{drawColor}{gray}{0.30}

\node[text=drawColor,anchor=base,inner sep=0pt, outer sep=0pt, scale=  0.80] at ( 37.94, 17.65) {0};

\node[text=drawColor,anchor=base,inner sep=0pt, outer sep=0pt, scale=  0.80] at ( 79.68, 17.65) {5};

\node[text=drawColor,anchor=base,inner sep=0pt, outer sep=0pt, scale=  0.80] at (121.42, 17.65) {10};

\node[text=drawColor,anchor=base,inner sep=0pt, outer sep=0pt, scale=  0.80] at (163.16, 17.65) {15};

\node[text=drawColor,anchor=base,inner sep=0pt, outer sep=0pt, scale=  0.80] at (204.90, 17.65) {20};
\end{scope}
\begin{scope}
\definecolor{drawColor}{RGB}{0,0,0}

\node[text=drawColor,anchor=base,inner sep=0pt, outer sep=0pt, scale=  0.80] at (130.82,  7.44) {Prevalence (\%)};
\end{scope}
\begin{scope}
\definecolor{drawColor}{RGB}{0,0,0}

\node[text=drawColor,rotate= 90.00,anchor=base,inner sep=0pt, outer sep=0pt, scale=  0.80] at ( 11.01, 74.54) {Experiment instance (sorted)};
\end{scope}
\begin{scope}
\definecolor{drawColor}{RGB}{211,211,211}
\definecolor{fillColor}{RGB}{255,255,255}

\path[draw=drawColor,line width= 0.6pt,line join=round,line cap=round,fill=fillColor] ( 93.89, 73.59) rectangle (229.05,121.92);
\end{scope}
\begin{scope}
\definecolor{drawColor}{RGB}{0,0,0}

\node[text=drawColor,anchor=base west,inner sep=0pt, outer sep=0pt, scale=  0.70] at ( 99.39,110.63) {Error Type};
\end{scope}
\begin{scope}
\definecolor{fillColor}{RGB}{12,44,132}

\path[fill=fillColor] (100.10,100.10) rectangle (104.70,105.44);
\end{scope}
\begin{scope}
\definecolor{fillColor}{RGB}{29,145,192}

\path[fill=fillColor] (100.10, 93.34) rectangle (104.70, 98.68);
\end{scope}
\begin{scope}
\definecolor{fillColor}{RGB}{127,205,187}

\path[fill=fillColor] (100.10, 86.57) rectangle (104.70, 91.91);
\end{scope}
\begin{scope}
\definecolor{fillColor}{RGB}{237,248,177}

\path[fill=fillColor] (100.10, 79.81) rectangle (104.70, 85.15);
\end{scope}
\begin{scope}
\definecolor{drawColor}{RGB}{0,0,0}

\node[text=drawColor,anchor=base west,inner sep=0pt, outer sep=0pt, scale=  0.70] at (108.92,100.36) {Dropped frames (Preprocessing)};
\end{scope}
\begin{scope}
\definecolor{drawColor}{RGB}{0,0,0}

\node[text=drawColor,anchor=base west,inner sep=0pt, outer sep=0pt, scale=  0.70] at (108.92, 93.60) {Dropped frames (Computer Vision)};
\end{scope}
\begin{scope}
\definecolor{drawColor}{RGB}{0,0,0}

\node[text=drawColor,anchor=base west,inner sep=0pt, outer sep=0pt, scale=  0.70] at (108.92, 86.83) {Input mismatches (Computer Vision)};
\end{scope}
\begin{scope}
\definecolor{drawColor}{RGB}{0,0,0}

\node[text=drawColor,anchor=base west,inner sep=0pt, outer sep=0pt, scale=  0.70] at (108.92, 80.07) {Dropped vehicles (EBA)};
\end{scope}
\end{tikzpicture}

%% file: 50_related_work.tex
\section{Related Work}
\label{sec:related-work} We are unaware of any publications that address
determinism in AUTOSAR \ac{AP}. In AUTOSAR~\ac{CP}, deterministic execution can
be achieved based on the \ac{LET}
paradigm~\cite{kirsch2012-logical-execution,autosar2018-specific-timing-extension,biondi2018-achievin-predictab},
which can also be extended for use in distributed
applications~\cite{ernst2018-system-level-let}.  However, while \ac{LET} is
compatible with the task-based programming model of AUTOSAR~\ac{CP}, it is not
easily applicable to the adaptive and reactive programming required in
AUTOSAR~\ac{AP}.  In particular, \ac{LET} is a real-time programming paradigm
where logical time strictly matches physical time at each task invocation and
termination.  The reactor semantics, in contrast, provides for a more flexible
application design where real-time constraints are only enforced when this is
explicitly indicated by a deadline. Time violations are not treated as system
failure but become observable errors that can be handled as appropriate for the
given application.  Another difference is that, while \ac{LET} tasks always
take a non-zero amount of logical time, reactions are logically instantaneous
and thus compose without requiring explicit alignment.

Synchronous languages like LUSTRE~\cite{halbwachs1991synchronous},
Esterel~\cite{berry1992esterel}, and SCADE~\cite{scade-berry} have been
designed for the development of complex reactive systems and are completely
deterministic. While the semantics of reactors is also a synchronous one,
reactors provide an additional coupling to physical time that allows for the
specification of real-time requirements and can be used for deterministic
distributed execution without a central coordinator.

%% file: 60_conclusion.tex
\section{Conclusion}
\label{sec:conclusion}

We have shown that AUTOSAR \ac{AP} exhibits nondeterminism in the core elements
of its architecture.  Our case study of a brake assistant demonstrator
application exposes that this could lead to serious malfunction.  To address
this, we propose to use reactors as programming model for the design of
applications in AUTOSAR \ac{AP}. We introduce the DEAR framework that
effectively enables reactor-based programming and coordination between
distributed components as shown in our case study. While our approach warrants
standard compatibility, we advocate for an extension of the standard that
obviates the need for the workarounds we implemented to associate method calls
and events with tags.